# A Primer on 3GPP Narrowband Internet of Things (NB-IoT)

Y.-P. Eric Wang[1], Xingqin Lin[1], Ansuman Adhikary[1], Asbjörn Grövlen[2], Yutao Sui[2], Yufei Blankenship[2], Johan Bergman[2], and Hazhir S. Razaghi[1]

Ericsson Research[1], Ericsson AB[2]

{eric.yp.wang, xingqin.lin, ansuman.adhikary, asbjorn.grovlen, yutao.sui, yufei.blankenship, johan.bergman, hazhir.shokri.razaghi}@ericsson.com

*Abstract*—**Narrowband Internet of Things (NB-IoT) is a new cellular technology introduced in 3GPP Release 13 for providing wide-area coverage for the Internet of Things (IoT). This article provides an overview of the air interface of NB-IoT. We describe how NB-IoT addresses key IoT requirements such as deployment flexibility, low device complexity, long battery life time, support of massive number of devices in a cell, and significant coverage extension beyond existing cellular technologies. We also share the various design rationales during the standardization of NB-IoT in Release 13 and point out several open areas for future evolution of NB-IoT.**

## I. INTRODUCTION

Use cases for machine-type communications are developing very rapidly. There has been enormous interest in integrating connectivity solutions with sensors, actuators, meters (water, gas, electric, or parking), cars, appliances, etc. [1,2]. The Internet of Things (IoT) is thus being created and constantly expanded. IoT consists of a number of networks that may have different design objectives. For example, some networks only intend to cover local area (e.g. one single home) whereas some networks offer wide-area coverage. The latter case is being addressed in the 3rd Generation Partnership Project (3GPP). Recognizing the importance of IoT, 3GPP has introduced a number of key features for IoT in its latest release, Rel-13. EC-GSM-IoT [3] and LTE-MTC [4] aim to enhance existing Global System for Mobile Communications (GSM) [5] and Long-Term Evolution (LTE) [6] networks, respectively, for better serving IoT use cases. Coverage extension, UE complexity reduction, long battery lifetime, and backward compatibility are common objectives. A third track, Narrowband Internet of Things (NB-IoT) [7], shares these objectives as well. In addition, NB-IoT aims to offer deployment flexibility allowing an operator to introduce NB-IoT using a small portion of its existing available spectrum. NB-IoT is designed mainly targeting ultra-low-end IoT applications.

NB-IoT is a new 3GPP radio-access technology in the sense that it is not fully backward compatible with existing 3GPP devices. It is however designed to achieve excellent co-existence performance with legacy GSM, General Packet Radio Service (GPRS) and LTE technologies. NB-IoT requires 180 kHz minimum system bandwidth for both downlink and uplink, respectively. The choice of minimum system bandwidth enables a number of deployment options. A GSM operator can replace one GSM carrier (200 kHz) with NB-IoT. An LTE operator can deploy NB-IoT inside an LTE carrier by allocating one of the Physical Resource Blocks (PRB) of 180 kHz to NB-IoT. As will become clear later in this article, the air interface of NB-IoT is optimized to ensure harmonious coexistence with LTE, and thus such an "in-band" deployment of NB-IoT inside an LTE carrier will not compromise the performance of LTE or NB-IoT. An LTE operator also has the option of deploying NB-IoT in the guard-band of the LTE carrier.

NB-IoT reuses the LTE design extensively, including the numerologies, downlink orthogonal frequency-division multiple-access (OFDMA), uplink single-carrier frequency-division multiple-access (SC-FDMA), channel coding, rate matching, interleaving, etc. This significantly reduces the time required to develop full specifications. Also, it is expected that the time required for developing NB-IoT products will be significantly reduced for existing LTE equipment and software vendors. The normative phase of NB-IoT work item in 3GPP started in September 2015 [7] and the core specifications complete in June 2016. Commercial launch of NB-IoT products and services is expected to be around the end of 2016 and the beginning of 2017.

In this article, we provide a state-of-the-art overview of the air interface of NB-IoT with a focus on the key aspects where NB-IoT deviates from LTE. In particular, we highlight the NB-IoT features that help achieve the aforementioned design objectives. The remainder of this article is organized as follows. In section II, transmission schemes and deployment options are given. Section III describes the physical channels of NB-IoT. Section IV describes resource mapping with an emphasis on how orthogonality with LTE is achieved when deploying NB-IoT inside an LTE carrier. Procedures such as cell search, random access, scheduling and hybrid automatic repeat request (HARQ) are detailed in sections V, VI, and VII, respectively. Section VIII highlights NB-IoT performance and section IX provides a conclusion.

## II. Transmission Schemes and Deployment Options

### A. Downlink transmission scheme

The downlink of NB-IoT is based on OFDMA with the same 15 kHz subcarrier spacing as LTE. Slot, subframe, and frame durations are 0.5 ms, 1 ms, and 10 ms, respectively, identical to those in LTE. Furthermore, slot format in terms of cyclic prefix (CP) duration and number of OFDM symbols per slot are also identical to those in LTE. In essence, an NB-IoT carrier uses one LTE PRB in the frequency domain, i.e. twelve 15 kHz subcarriers for a total of 180 kHz. Reusing the same OFDM numerology as LTE ensures the coexistence performance with LTE in the downlink. For example, when NB-IoT is deployed inside an LTE carrier, the orthogonality between the NB-IoT PRB and all the other LTE PRBs is preserved in the downlink.

### B. Uplink transmission scheme

The uplink of NB-IoT supports both multi-tone and single-tone transmissions. Multi-tone transmission is based on SC-FDMA with the same 15 kHz subcarrier spacing, 0.5 ms slot, and 1 ms subframe as LTE. Single-tone transmission supports two numerologies, 15 kHz and 3.75 kHz. The 15 kHz numerology is identical to LTE and thus achieves the best coexistence performance with LTE in the uplink. The 3.75 kHz single-tone numerology uses 2 ms slot duration. Like the downlink, an uplink NB-IoT carrier uses a total system bandwidth of 180 kHz.

### C. Deployment options

NB-IoT may be deployed as a stand-alone carrier using any available spectrum exceeding 180 kHz. It may also be deployed within the LTE spectrum allocation, either inside an LTE carrier or in the guard band. These different deployment scenarios are illustrated in Fig. 1. The deployment scenario, stand-alone, in-band, or guard-band, however should be transparent to a user equipment (UE) when it is first turned on and searches for an NB-IoT carrier. Similar to existing LTE UEs, an NB-IoT UE is only required to search for a carrier on a 100 kHz raster. An NB-IoT carrier that is intended for facilitating UE initial synchronization is referred to as an *anchor* carrier. The 100 kHz UE search raster implies that for in-band deployments, an anchor carrier can only be placed in certain PRBs. For example, in a 10 MHz LTE carrier, the indexes of the PRBs that are best aligned with the 100 kHz grid and can be used as an NB-IoT anchor carrier are 4, 9, 14, 19, 30, 35, 40, 45. The PRB indexing starts from index 0 for the PRB occupying the lowest frequency within the LTE system bandwidth.

Fig. 1 illustrates the deployment options of NB-IoT with a 10 MHz LTE carrier. The PRB right above the DC subcarrier, i.e., PRB #25, is centered at 97.5 kHz (i.e. a spacing of 6.5 subcarriers) above the DC subcarrier. Since the LTE DC subcarrier is placed on the 100 kHz raster, the center of PRB#25 is 2.5 kHz from the nearest 100 kHz grid. The spacing between the centers of two neighboring PRBs above the DC subcarrier is 180 kHz. Thus, PRB #30, #35, #40, and #45 are all centered at 2.5 kHz from the nearest 100 kHz grid. It can be shown that for LTE carriers of 10 MHz and 20 MHz, there exists a set of PRB indexes that are all centered at 2.5 kHz from the nearest 100 kHz grid, whereas for LTE carriers of 3 MHz, 5 MHz, and 15 MHz bandwidth, the PRB indexes are centered at least 7.5 kHz away from the 100 kHz raster. Further, an NB-IoT anchor carrier should not be any of the middle 6 PRBs of the LTE carrier (e.g. PRB#25 of 10 MHz LTE, although its center is 2.5 kHz from the nearest 100 kHz raster). This is due to that LTE synchronization and broadcast channels occupy many resource elements in the middle 6 PRBs, making it difficult to use these PRBs for NB-IoT.

Similar to the in-band deployment, an NB-IoT anchor carrier in the guard-band deployment needs to have center frequency no more than 7.5 kHz from the 100 kHz raster. NB-IoT cell search and initial acquisition are designed for a UE to be able to synchronize to the network in the presence of a raster offset up to 7.5 kHz.

Multi-carrier operation of NB-IoT is supported. Since it suffices to have one NB-IoT anchor carrier for facilitating UE initial synchronization, the additional carriers do not need to be near the 100 kHz raster grid. These additional carriers are referred to as secondary carriers.

## III. Physical Channels

NB-IoT physical channels are designed based on legacy LTE to a large extent. In this section, we provide an overview of them with a focus on aspects that are different from legacy LTE.

### A. Downlink

NB-IoT provides the following physical signals and

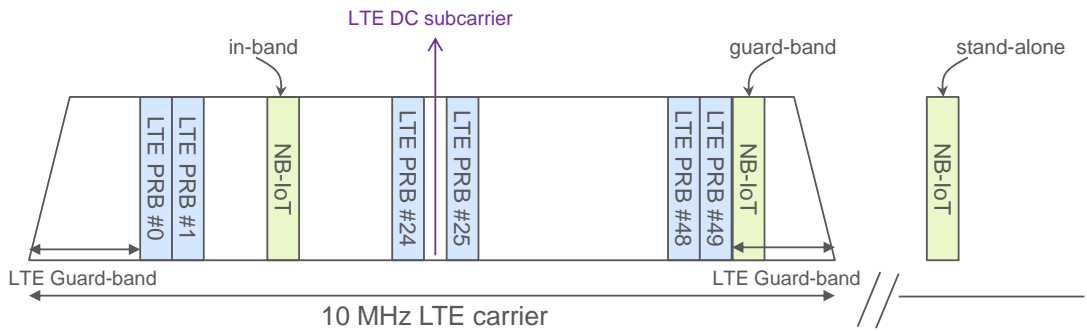

Fig. 1. Examples of NB-IoT stand-alone deployment and LTE in-band and guard-band deployments in the downlink.

| | subframe number | | | | | | | | | |
|---|---|---|---|---|---|---|---|---|---|---|
| even numbered frame | 0 | 1 | 2 | 3 | 4 | 5 | 6 | 7 | 8 | 9 |
| | NPBCH | NPDCCH or NPDSCH | NPDCCH or NPDSCH | NPDCCH or NPDSCH | NPDCCH or NPDSCH | NPSS | NPDCCH or NPDSCH | NPDCCH or NPDSCH | NPDCCH or NPDSCH | NSSS |
| | subframe number | | | | | | | | | |
| odd numbered frame | 0 | 1 | 2 | 3 | 4 | 5 | 6 | 7 | 8 | 9 |
| | NPBCH | NPDCCH or NPDSCH | NPDCCH or NPDSCH | NPDCCH or NPDSCH | NPDCCH or NPDSCH | NPSS | NPDCCH or NPDSCH | NPDCCH or NPDSCH | NPDCCH or NPDSCH | NPDCCH or NPDSCH |

Fig. 2. Time multiplexing between NB-IoT downlink physical channels and signals.

channels in the downlink.
- Narrowband Primary Synchronization Signal (NPSS)
- Narrowband Secondary Synchronization Signal (NSSS)
- Narrowband Physical Broadcast Channel (NPBCH)
- Narrowband Reference Signal (NRS)
- Narrowband Physical Downlink Control Channel (NPDCCH)
- Narrowband Physical Downlink Shared Channel (NPDSCH)

Unlike LTE, these NB-IoT physical channels and signals are primarily multiplexed in time. Fig. 2 illustrates how the NB-IoT subframes are allocated to different physical channels and signals. Each NB-IoT subframe spans over one PRB (i.e. 12 subcarriers) in the frequency domain and 1 ms in the time domain.

NPSS and NSSS are used by an NB-IoT UE to perform cell search, which includes time and frequency synchronization, and cell identity detection. Since the legacy LTE synchronization sequences occupy 6 PRBs, they cannot be reused for NB-IoT. A new design is thus introduced.

NPSS is transmitted in subframe #5 in every 10 ms frame, using the last 11 OFDM symbols in the subframe. NPSS detection is one of the most computationally demanding operations from a UE perspective. To allow efficient implementation of NPSS detection, NB-IoT uses a hierarchical sequence. For each of the 11 NPSS OFDM symbols in a subframe, either $p$ or $-p$ is transmitted, where $p$ is the base sequence generated based on a length-11 Zadoff-Chu (ZC) sequence [8] with root index 5. Each of the length-11 ZC sequence is mapped to the lowest 11 subcarriers within the NB-IoT PRB.

NSSS has 20 ms periodicity and is transmitted in subframe #9, also using the last 11 OFDM symbols that consist of 132 resource elements overall. NSSS is a length-132 frequency-domain sequence, with each element mapped to a resource element. NSSS is generated by element-wise multiplication between a ZC sequence and a binary scrambling sequence. The root of the ZC sequence and binary scrambling sequence are determined by narrowband physical cell identity (NB-PCID). The cyclic shift of the ZC sequence is further determined by the frame number.

NPBCH carries the master information block (MIB) and is transmitted in subframe #0 in every frame. A MIB remains unchanged over the 640 ms transmission time interval (TTI).

NPDCCH carries scheduling information for both downlink and uplink data channels. It further carries the HARQ acknowledgement information for the uplink data channel as well as paging indication and random access response (RAR) scheduling information. NPDSCH carries data from the higher layers as well as paging message, system information, and the RAR message. As shown in Fig. 2, there are a number of subframes that can be allocated to carry NPDCCH or NPDSCH. To reduce UE complexity, all the downlink channels use the LTE tail-biting convolutional code (TBCC). Furthermore, the maximum transport block size of NPDSCH is 680 bits. In comparison, LTE without spatial multiplexing supports maximum TBS greater than 70,000 bits.

NRS is used to provide phase reference for the demodulation of the downlink channels. NRSs are time-and-frequency multiplexed with information bearing symbols in subframes carrying NPBCH, NPDCCH and NPDSCH, using 8 resource elements per subframe per antenna port.

*B. Uplink*

NB-IoT includes the following channels in the uplink.
- Narrowband Physical Random Access Channel (NPRACH)
- Narrowband Physical Uplink Shared Channel (NPUSCH)

NPRACH is a newly designed channel since the legacy LTE Physical Random Access Channel (PRACH) uses a bandwidth of 1.08 MHz, more than NB-IoT uplink bandwidth. One NPRACH preamble consists of 4 symbol groups, with each symbol group comprising of one CP and 5 symbols. The CP length is 66.67 μs (Format 0) for cell radius up to 10 km and 266.7 μs (Format 1) for cell radius up to 40 km. Each symbol, with fixed symbol value 1, is modulated on a 3.75 kHz tone with symbol duration of 266.67 μs. However, the tone frequency index changes from one symbol group to another. The waveform of NPRACH preamble is referred to as single-tone frequency hopping. An example of NPRACH frequency hopping is illustrated in Fig. 3. To support coverage extension, a NPRACH preamble can be repeated up to 128 times.

NPUSCH has two formats. Format 1 is used for carrying uplink data and uses the same LTE turbo code for error correction. The maximum transport block size of NPUSCH Format 1 is 1000 bits, which is much lower than that in LTE. Format 2 is used for signaling HARQ acknowledgement for NPDSCH, and uses a repetition code for error correction. NPUSCH Format 1 supports multi-tone transmission based on the same legacy LTE numerology. In this case, the UE can be allocated with 12, 6, or 3 tones. While only the 12-tone format is supported by legacy LTE UEs, the 6-tone and 3-tone formats are introduced for NB-IoT UEs who due to coverage limitation cannot benefit from higher UE bandwidth allocation. Moreover, NPUSCH supports single-tone transmission based on either 15 kHz or 3.75 kHz numerology. To reduce peak-to-average power ratio (PAPR), single-tone transmission uses $\pi/2$-BPSK or $\pi/4$-QPSK with phase continuity between symbols.

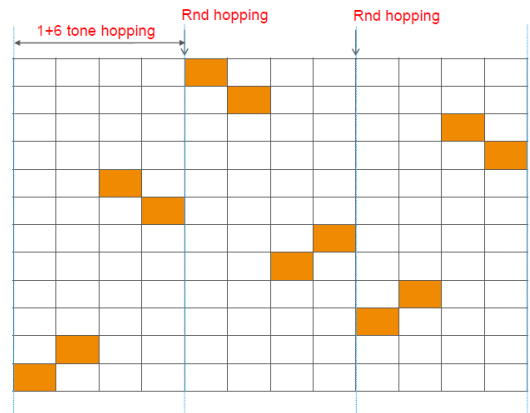

Fig. 3. An illustration of NPRACH frequency hopping.

NPUSCH Format 1 uses the same slot structure as legacy LTE PUSCH with 7 OFDM symbols per slot and the middle symbol as the demodulation reference symbol (DMRS). NPUSCH Format 2 also has 7 OFDM symbols per slot, but uses the middle three symbols as DMRS. DMRS are used for channel estimation.

Table 1 summarizes the NB-IoT physical channels and their differences with the LTE counterparts.

## IV. RESOURCE MAPPING

In this section, we describe how NB-IoT resource mapping is designed to ensure the best coexistence performance with LTE if deployed inside an LTE carrier. In essence, the orthogonality to LTE signals is preserved by avoiding mapping NB-IoT signals to the resource elements already used by the legacy LTE signals. An example is illustrated in Fig. 4, in which each column indicates resource elements in one OFDM symbol. There are 12 resource elements per OFDM symbol corresponding to 12 subcarriers. As shown, for the stand-alone and guard-band deployments, no LTE resource needs to be protected, thus NPDCCH, NPDSCH or NRS can utilize all the resource elements in one PRB pair (defined as 12 subcarriers over one subframe). However, for in-band deployment, NPDCCH, NPDSCH or NRS cannot be mapped to the resource elements taken by LTE Cell-Specific Reference Symbols (CRS) and LTE Physical Downlink Control Channel (PDCCH). NB-IoT is designed to allow a UE to learn the deployment mode (stand-alone, in-band, or guard-band) as well as the cell identity (both NB-IoT and LTE) through initial acquisition. Then the UE can figure out which resource elements are used by LTE. With this information, the UE can map NPDCCH and NPDSCH symbols to available resource elements. On the other hand, NPSS, NSSS, and NPBCH are used for initial synchronization and master system information acquisition. These signals need to be detected without knowing the deployment mode. To facilitate this, NPSS, NSSS, and NPBCH avoid the first three OFDM symbols in every subframe as these resource elements may be used by LTE PDCCH. Furthermore, NPSS and NSSS signals overlapping with resource elements taken by LTE CRS are punctured at the base station. Although the UE is not aware of which resource elements are punctured, NPSS and NSSS can still be detected by correlating the received punctured NPSS and NSSS signal with the non-punctured signal since the percentage of punctured resource elements is relatively small. NPBCH is rate-matched around LTE CRS. This however requires the UE to figure out the location of CRS resource elements, which is dependent of LTE physical cell identity (PCID). Recall that the UE learns cell identity (NB-PCID) from NSSS. The relationship of the values of PCID and NB-PCID used by the same cell is such that the UE can use NB-PCID to determine the LTE CRS locations.

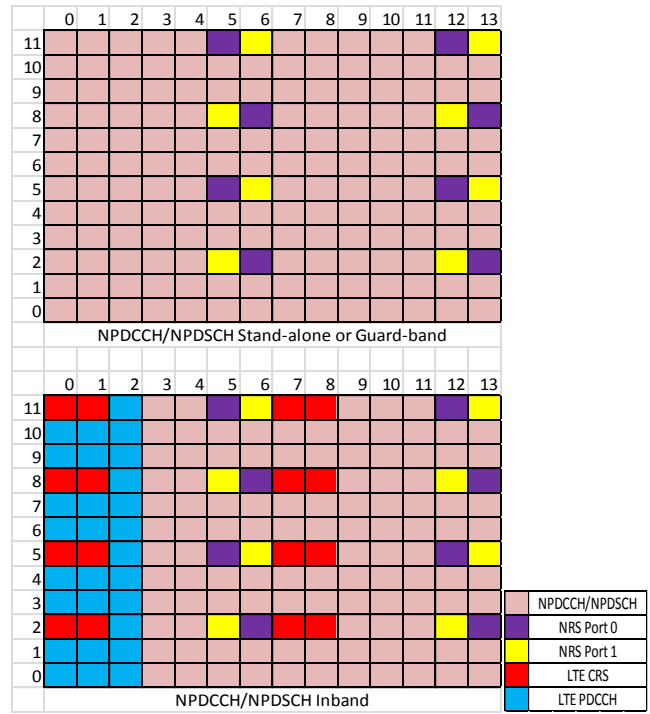

Fig. 4. NPDCCH/NPDSCH resource mapping example.

Table 1. Summary of NB-IoT physical signals and channels and their relationship with the LTE counterparts.

| | Physical Channel | Relationship with LTE |
|---|---|---|
| Downlink | NPSS | • New sequence for fitting into one PRB (LTE PSS overlaps with middle 6 PRBs)<br>• All cells share one NPSS (LTE uses 3 PSSs) |
| | NSSS | • New sequence for fitting into one PRB (LTE SSS overlaps with middle 6 PRBs)<br>• NSSS provides the lowest 3 least significant bits of system frame number (LTE SSS does not) |
| | NPBCH | • 640 ms TTI (LTE uses 40 ms TTI) |
| | NPDCCH | • May use multiple PRBs in time, i.e. multiple subframes (LTE PDCCH uses multiple PRBs in frequency and 1 subframe in time) |
| | NPDSCH | • Use TBCC and only one redundancy version (LTE uses Turbo Code with multiple redundancy versions)<br>• Use only QPSK (LTE also uses higher order modulations)<br>• Maximum transport block size (TBS) is 680 bits. (LTE without spatial multiplexing has maximum TBS greater than 70000 bits, see [9])<br>• Supports only single-layer transmission (LTE can support multiple spatial-multiplexing layers) |
| Uplink | NPRACH | • New preamble format based on single-tone frequency hopping using 3.75 kHz tone spacing (LTE PRACH occupies 6 PRBs and uses multi-tone transmission format with 1.25 kHz subcarrier spacing) |
| | NPUSCH Format 1 | • Support UE bandwidth allocation smaller than one PRB (LTE has minimum bandwidth allocation of 1 PRB)<br>• Support both 15 kHz and 3.75 kHz numerology for single-tone transmission (LTE only uses 15 kHz numerology)<br>• Use $\pi/2$-BPSK or $\pi/4$-QPSK for single-tone transmission (LTE uses regular QPSK and higher order modulations)<br>• Maximum TBS is 1000 bits. (LTE without spatial multiplexing has maximum TBS greater than 70000 bits, see [9])<br>• Supports only single-layer transmission (LTE can support multiple spatial-multiplexing layers) |
| | NPUSCH Format 2 | • New coding scheme (repetition code)<br>• Use only single-tone transmission |

## V. Cell Search and Initial Acquisition Procedure

Synchronization is an important aspect in cellular communications. When a UE is powered on for the first time, it needs to detect a suitable cell to camp on, and for that cell, obtain the symbol, subframe, and frame timing as well as synchronize to the carrier frequency. In order to synchronize to the carrier frequency, the UE needs to correct any erroneous frequency offsets that are present due to local oscillator inaccuracy, and perform symbol timing alignment with the frame structure from the base station. In addition, due to the presence of multiple cells, the UE needs to distinguish a particular cell on the basis of an NB-PCID. As a result, a typical synchronization procedure consists of determining the timing alignment, correcting the frequency offset, obtaining the correct cell identity, and the absolute subframe and frame number reference.

NB-IoT is intended to be used for very low cost UEs and at the same time, provide extended coverage for UEs deployed in environments with high penetration losses, e.g., basement of a building. Such low cost UEs are equipped with low-cost crystal oscillators that can have an initial carrier frequency offset (CFO) as large as 20 ppm. Deployment in-band and in guard-bands of LTE introduces an additional raster offset (2.5 or 7.5 kHz) as explained in Section II, giving rise to an even higher CFO. Despite of this large CFO, a UE should also be able to perform accurate synchronization at very low SNR.

Synchronization in NB-IoT follows similar principles as the synchronization process in LTE, but with changes to the design of the synchronization sequences in order to resolve the problem of estimating large frequency offset and symbol timing at very low SNR. Synchronization is achieved through the use of NPSS and NSSS. As mentioned in Section III, the NPSS occurs in subframe #5 of every frame, and the NSSS occurs in subframe #9 of every even numbered frame. The NPSS is used to obtain the symbol timing and the CFO, and the NSSS is used to obtain the NB-PCID, and the timing within an 80 ms block.

For UEs operating at very low SNR, an auto correlation based on a single 10 ms received segment would not be sufficient for detection. As a result, an accumulation procedure over multiple 10 ms segments is necessary. Because of the inherent NPSS design, the accumulation can be performed coherently, providing sufficient signal energy for detection. Because of the large initial CFO, the sampling time at the UE is different from the actual sampling time, the difference being proportional to the CFO. For UEs in deep coverage, the number of accumulations necessary to achieve a successful detection may be high. As a result, the peak of every accumulation process does not add up coherently because of the difference in the true and UE sampling time causing a drift. The drift can be handled by using a weighted accumulation procedure, so that the most recent accumulated value is given higher priority than the previous ones.

After the synchronization procedure is complete, the UE has knowledge of the symbol timing, the CFO, the position within an 80 ms block and the NB-PCID. The UE then proceeds to the acquisition of the MIB, which is broadcast in subframe #0 of every frame carried by NPBCH. The NPBCH consists of 8 self-decodable sub-blocks, and each sub-block is repeated 8 times so that each sub-block occupies subframe #0 of 8 consecutive frames. The design is intended to provide successful acquisition for UEs in deep coverage.

After the symbol timing is known and the CFO is compensated for, in the in-band and guard-band deployment there is still an additional raster offset which can be as high as 7.5 kHz. The presence of raster offset results in either overcompensation or under compensation of the carrier frequency. As a result, the symbol timing drifts in either the forward or backward direction depending on whether the carrier frequency was overcompensated or undercompensated. This may cause a severe degradation in the performance of NPBCH detection if the NPBCH is not detected on the first trial. For example, an unsuccessful detection of NPBCH in the first trial introduces a latency of 640 ms before the next NPBCH detection trial. A 7.5 kHz raster offset leads to a symbol timing drift of 5.33 μs (assuming a carrier frequency of 900 MHz) which is greater than the duration of cyclic prefix. As a result, the downlink orthogonality of OFDM is lost. A solution to this problem comes at the expense of a small increase in computational complexity, where the UE can perform a "hypothesis testing" over the set of possible raster offsets to improve the detection performance. Since the number of possible raster offsets is small and there is only one NPBCH subframes in every 10 subframes, this is feasible from an implementation point of view.

## VI. Random Access

In NB-IoT, random access serves multiple purposes such as initial access when establishing a radio link and scheduling request. Among others, one main objective of random access is to achieve uplink synchronization, which is important for maintaining uplink orthogonality in NB-IoT. Similar to LTE, the contention-based random access procedure in NB-IoT consists of four steps: (1) UE transmits a random access preamble; (2) the network transmits a random access response that contains timing advance command and scheduling of uplink resources for the UE to use in the third step; (3) the UE transmits its identity to the network using the scheduled resources; and (4) the network transmits a contention-resolution message to resolve any contention due to multiple UEs transmitting the same random access preamble in the first step.

To serve UEs in different coverage classes that have different ranges of path loss, the network can configure up to three NPRACH resource configurations in a cell. In each configuration, a repetition value is specified for repeating a basic random access preamble. UE measures its downlink received signal power to estimate its coverage level, and transmits random access preamble in the NPRACH resources configured for its estimated coverage level. To facilitate NB-IoT deployment in different scenarios, NB-IoT allows flexible configuration of NPRACH resources in time-frequency resource grid with the following parameters.

- Time domain: periodicity of NPRACH resource, and starting time of NPRACH resource in a period.
- Frequency domain: frequency location (in terms of subcarrier offset), and number of subcarriers.

It is possible that in the early NB-IoT field trial and deployment, some UE implementations may not support multi-tone transmission. The network should be aware of UE multi-tone transmission capability before scheduling uplink transmission. Therefore, the UE should indicate its support of multi-tone transmission in the first step of random access to facilitate the network's scheduling of uplink transmission in the third step of random access. To this end, the network can partition the NPRACH subcarriers in the frequency domain into two non-overlapping sets. A UE can select one of the two sets to transmit its random access preamble to signal whether or not it supports multi-tone transmission in the third step of random access.

In summary, UE determines its coverage level by measuring downlink received signal power. After reading system information on NPRACH resource configuration, the UE can determine the NPRACH resource configured and the numbers of repetitions needed for its estimated coverage level as well as random access preamble transmit power. Then the UE can transmit the repetitions of the basic single tone random access preamble back-to-back within one period of the NPRACH resources. The remaining steps in random access procedure are similar to LTE, and we omit the details here.

## VII. Scheduling and HARQ Operation

To enable low-complexity UE implementation, NB-IoT allows only one HARQ process in both downlink and uplink, and allows longer UE decoding time for both NPDCCH and NPDSCH. Asynchronous, adaptive HARQ procedure is adopted to support scheduling flexibility. An example is illustrated in Fig. 5. Scheduling command is conveyed through Downlink Control Indicator (DCI), which is carried by NPDCCH. NPDCCH may use aggregation levels (AL) 1 or 2 for transmitting a DCI. With AL-1, two DCIs are multiplexed in one subframe, otherwise one subframe only carries one DCI (i.e. AL-2), giving rise to a lower coding rate and improved coverage. Further coverage enhancement can be achieved through repetition. Each repetition occupies one subframe. DCI can be used for scheduling downlink data or uplink data. In the case of downlink data, the exact time offset between NPDCCH and the associated NPDSCH is indicated in the DCI. Since IoT devices are expected to have reduced computing capability, the time offset between the end of NPDCCH and the beginning of the associated NPDSCH is at least 4 ms. In comparison, LTE PDCCH schedules PDSCH in the same TTI. After receiving NPDSCH, the UE needs to send back HARQ acknowledgement using NPUSCH Format 2. The resources of NPUSCH carrying HARQ acknowledgement are also indicated in DCI. Considering the limited computing resources in an IoT device, the time offset between the end of NPDSCH and the start of the associated HARQ acknowledgement is at least 12 ms. This offset is longer than that between NPDCCH and NPDSCH because the transport block carried in NPDSCH might be up to 680 bits, a lot longer than the DCI, which is only 23 bits long.

Similarly, uplink scheduling and HARQ operation are also illustrated in Fig. 5. The DCI for uplink scheduling grant needs to specify which subcarriers that a UE is allocated. The time offset between the end of NPDCCH and the beginning of the associated NPUSCH is at least 8 ms. After completing the NPUSCH transmission, the UE monitors NPDCCH to learn whether NPUSCH is received correctly by the base station, or a retransmission is needed.

## VIII. Performance

IoT use cases are characterized by requirements such as data rate, coverage, device complexity, latency, and battery lifetime. These are thus important performance metrics. Furthermore, according to [10], IoT traffic is forecasted to have compounded annual growth rate of 23% between 2015 and 2023. It is therefore important to ensure that NB-IoT has good capacity to support such a growth in the years to come. In this section, we discuss NB-IoT performance in the aspects mentioned above.

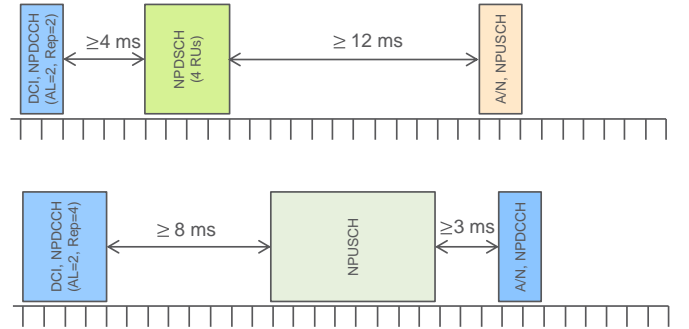

Fig. 5. Timing relationship operation (each unit corresponds to one subframe)

*A. Peak Data Rates*

NDSCH peak data rate can be achieved by using the largest TBS of 680 bits and transmitting it over 3 ms. This gives 226.7 kbps peak layer-1 data rate. NPUSCH peak data rate can be achieved by using the largest TBS of 1000 bits and transmitting it over 4 ms. This gives 250 kbps peak layer-1 data rate. However, the peak throughputs of both downlink and uplink are lower than the above figures, when the time offsets between DCI, NPDSCH/NPUSCH, and HARQ acknowledgement are taken into account.

*B. Coverage*

NB-IoT achieves a maximum coupling loss 20 dB higher than LTE Rel-12 [11, 12]. Coverage extension is achieved by trading off data rate through increasing the number of repetitions. Coverage enhancement is ensured also by introducing single subcarrier NPUSCH transmission and $\pi/2$-BPSK modulation to maintain close to 0 dB PAPR, thereby reducing the unrealized coverage potential due to power amplifier (PA) backoff. NPUSCH with 15 kHz single-tone gives a layer-1 data rate of approximately 20 bps when configured with the highest repetition factor, i.e., 128, and the lowest modulation and coding scheme. NPDSCH gives a layer-1 data rate of 35 bps when configured with repetition factor 512 and the lowest modulation and coding scheme. These configurations support close to 170 dB coupling loss. In comparison, the Rel-12 LTE network is designed for up to approximately 142 dB coupling loss [13].

*C. Device complexity*

NB-IoT enables low-complexity UE implementation by the designs highlighted below.

- Significantly reduced transport block sizes for both downlink and uplink
- Support only one redundancy version in the downlink
- Support only single-stream transmissions in both downlink and uplink
- A UE only requires single antenna
- Support only single HARQ process in both downlink and uplink
- No need for a turbo decoder at the UE since only TBCC is used for downlink channels
- No Connected mode mobility measurement is required. A UE only needs to perform mobility measurement during the Idle mode
- Low sampling rate due to lower UE bandwidth
- Allow only half-duplex frequency-division duplexing (FDD) operation
- No parallel processing is required. All the physical layer procedures and transmission and reception of physical channels occur in sequential manner

The coverage objective is achieved with 20 or 23 dBm PA, making it possible to use an integrated PA in the UE.

*D. Latency and battery lifetime*

NB-IoT targets latency insensitive applications. However, for applications like sending alarm signals, NB-IoT is designed to allow less than 10 s latency [3]. NB-IoT aims to support long battery life. For a device with 164 dB coupling loss, a 10-year battery life can be reached if the UE transmits 200-byte data a day on average [3].

*E. Capacity*

NB-IoT supports massive IoT capacity by using only one PRB in both uplink and downlink. Sub-PRB UE scheduled bandwidth is introduced in the uplink, including single subcarrier NPUSCH. Note that for coverage limited UE, allocating higher bandwidth is not spectrally efficient as the UE cannot benefit from it to be able to transmit at a higher data rate. Based on the traffic model in [3], NB-IoT with one PRB supports more than 52500 UEs per cell [3]. Furthermore, NB-IoT supports multiple carrier operation. Thus, more IoT capacity can be added by adding more NB-IoT carriers.

IX. CONCLUSION

In this article, a description of NB-IoT radio access is given. We emphasize on how radio access is designed differently compared to LTE, and how it is designed to fulfill the performance requirements of IoT such as significant coverage extension, low device complexity, long battery lifetime, and supporting a massive number of IoT devices. NB-IoT is also designed to allow easy integration and sharing of radio resources with existing GSM and LTE networks. Further enhancements of NB-IoT in the next 3GPP release are being discussed [14, 15], including, for example, introducing low complexity multicast functionality, for rolling out firmware updates, and enhancing positioning accuracy, which is important to many IoT applications. NB-IoT is a step toward building the fifth-generation (5G) radio access technology intended for enabling new use cases like machine type communications. It is foreseen that NB-IoT will continue to evolve toward 5G requirements